\documentclass[journal,a4paper]{IEEEtran}
\usepackage{amsthm,amscd}
\usepackage{amssymb,amsmath,epsfig,latexsym}
\theoremstyle{definition}
\newtheorem{definition}{Definition}[section]

\newtheorem{Voorb}[definition]{Example}
\theoremstyle{plain}
\newtheorem{Gevolg}[definition]{Corollary}
\newtheorem{Lemma}[definition]{Lemma}
\newtheorem{Stel}[definition]{Theorem}

\newcommand{\ruimte}{\par\vspace{1ex}\noindent}

\begin{document}

\title{On minimality of convolutional ring encoders}

\author{Margreta~Kuijper and~Raquel~Pinto%
        \thanks{M. Kuijper is with the Department
of EE Engineering, University of Melbourne, VIC 3010, Australia \texttt{E-mail: m.kuijper@ee.unimelb.edu.au}.}%
\thanks{R. Pinto is with the Department of Mathematics, University of Aveiro, 3810-193 Aveiro, Portugal \texttt{E-mail: raquel@ua.pt}.}}%

\newcommand{\ellL}{L}
\newcommand{\dubbel}[1]{{\mathbb #1}}
\newcommand{\F}{{\mathbb F}}
\newcommand{\Z}{{\mathbb Z}}
\newcommand{\Zpr}{{\mathbb Z}_{p^r}}
\newcommand{\es}{x}
\newcommand{\st}{ \;|\;}
\newcommand{\ruimtevier}{\par\vspace{4ex}\noindent}
\newcommand{\ruimtetien}{\par\vspace{10ex}\noindent}
\newcommand{\Fpolstimes}[2]{{\mathbb F}^{#1 \times #2} [s]}
\newcommand{\Rpolstimes}[2]{{\mathbb R}^{#1 \times #2} [s]}
\newcommand{\wt}{\tilde{\wbold}}
\newcommand{\et}{\tilde{\ebold}}
\newcommand{\suitc}{s \in \mC}
\newcommand{\keenN}{k = 1,2,\ldots , N}
\newcommand{\knulN}{k = 0,1,\ldots , N}
\newcommand{\arij}{a_1, a_2, \ldots ,a_N}
\newcommand{\ariji}{a_1^i, a_2^i, \ldots ,a_N^i}
\newcommand{\arijK}{a_1, a_2, \ldots ,a_K}
\newcommand{\Arijnul}{A_0, A_1, \ldots ,A_N}
\newcommand{\Arij}{A_1, \ldots ,A_N}
\newcommand{\arijk}{a_1, a_2, \ldots ,a_k}
\newcommand{\arijkp}{a_1, a_2, \ldots ,a_{k+1}}
\newcommand{\arijkm}{a_1, a_2, \ldots ,a_{k-1}}
\newcommand{\crij}{c_1, c_2, \ldots ,c_L}
\newcommand{\pfend}{\par\vspace{2ex}\noindent}
\newcommand{\eind}{\hspace*{\fill}$\Box$\par\vspace{2ex}\noindent}
\newcommand{\beq}{\begin{equation}}
\newcommand{\eeq}{\end{equation}}
\newcommand{\bmat}{\left[ \begin{array}}
\newcommand{\emat}{\end{array} \right]}
\newcommand{\twee}[2]{\left[ #1 \sfour #2 \right]}
\newcommand{\A}{{\cal A}}
\newcommand{\X}{{\cal X}}
\newcommand{\B}{{\cal B}}
\newcommand{\R}{{\cal R}}
\newcommand{\C}{{\cal C}}
\newcommand{\G}{{\cal G}}
\newcommand{\SUPP}{\mathrm{supp }\;}
\newcommand{\COL}{\mathrm{col }\;}
\newcommand{\ROW}{\mathrm{row }\;}
\newcommand{\PRANK}{p\mathrm{-rank }\;}
\newcommand{\PDIM}{p\mathrm{-dim }\;}
\newcommand{\PSPAN}{p\mathrm{-span }\;}
\newcommand{\SPAN}{\;\mathrm{span }\;}
\newcommand{\KER}{\mbox{ ker }}
\newcommand{\IM}{\mbox{ im }}
\newcommand{\AND}{\;\mbox{and }}
\newcommand{\WITH}{\;\mbox{with }}
\newcommand{\DET}{\mathrm{det }\;}
\newcommand{\DEG}{\mathrm{deg }\;}
\newcommand{\ROWDEG}{\mathrm{rowdeg }\;}
\newcommand{\PEXTDEG}{p\mathrm{-ext deg }\,}
\newcommand{\MAX}{\mathrm{max }\;}
\newcommand{\MOD}{\;\mbox{mod}}
\newcommand{\FOR}{\sfour\mbox{for }}
\newcommand{\IF}{\sfour\mbox{if }}
\newcommand{\OR}{\sfour\mbox{or }}
\newcommand{\la}{\lambda}
\newcommand{\punt}{ ,\cdots , }
\newcommand{\sfour}{ \;\;\;\; }
\newcommand{\vbold}{\mbox{\boldmath $v$}}
\newcommand{\wbold}{\mbox{\boldmath $w$}}
\newcommand{\wbbold}{\bar{\mbox{\boldmath $w$}}}
\newcommand{\abold}{\mbox{\boldmath $a$}}
\newcommand{\bbold}{\mbox{\boldmath $b$}}
\newcommand{\cbold}{\mbox{\boldmath $c$}}
\newcommand{\ctbold}{\tilde{\mbox{\boldmath $c$}}}
\newcommand{\cbarbold}{\bar{\mbox{\boldmath $c$}}}
\newcommand{\ibold}{\mbox{\boldmath $i$}}
\newcommand{\ebold}{\mbox{\boldmath $e$}}
\newcommand{\ybold}{\mbox{\boldmath $y$}}
\newcommand{\ubold}{\mbox{\boldmath $u$}}
\newcommand{\rbold}{\mbox{\boldmath $r$}}
\newcommand{\nlabel}[1]{\label{#1}}
\newcommand{\kopje}[1]{\ruimte \underline{\bf  #1}}

\maketitle

\ruimte
\begin{abstract}
Convolutional codes
are considered with code
sequences modelled as semi-infinite Laurent series.
It is wellknown that
a convolutional code $\C$ over a finite group ${\cal G}$ has a
minimal trellis representation that can be derived from code sequences.
It is also wellknown that, for the case that $\G$ is a finite field, any polynomial encoder of $\C$ can be algebraically manipulated to yield a minimal polynomial encoder whose controller canonical
realization is a minimal trellis. In this paper we seek to extend this result to the finite ring case ${\cal G} = \Zpr$ by introducing a
socalled ``$p$-encoder''.
We show how to manipulate a polynomial encoding of a noncatastrophic convolutional code over $\Zpr$ to produce a particular type of
$p$-encoder (``minimal $p$-encoder'') whose controller canonical
realization is a minimal trellis
with nonlinear features.
The minimum number of trellis states is then expressed as
$p^\gamma$, where $\gamma$ is the sum of the row degrees of the
minimal $p$-encoder. In particular, we show that any convolutional code over $\Zpr$ admits a delay-free $p$-encoder which implies the novel result that delay-freeness
is not a property of the code but of the encoder, just as in the field
case.
We conjecture that a similar result holds with respect to catastrophicity, i.e., any catastrophic
convolutional code over $\Zpr$ admits a noncatastrophic $p$-encoder.
\end{abstract}
\section{Introduction}\label{sec_intro}
There exists a considerable body of literature on
convolutional codes over
finite groups. In this paper we are
interested in trellis representations that use a minimum number of
states. Since decoders, such as the Viterbi decoder,
are based on trellis representations, minimality is a desirable
property that leads to low complexity decoding.
In~\cite[Sect.\ VI-D]{forneyT93} a minimal encoder construction is presented in terms of code sequences of the code, involving socalled ``granule representatives'', see also~\cite{loeligerM96}. This is a powerful method that applies to convolutional codes over any finite group $\G$. It is wellknown that, for the case that $\G$ is a field, any polynomial encoder of a convolutional code can be algebraically manipulated to yield a so-called ``canonical polynomial encoder'' (left prime and row reduced)
whose controller canonical realization yields
a minimal trellis representation of the code. This is a fundamental
result that is useful in practice because codes are usually specified
in terms of encoders rather than code sequences. In this paper we seek
to extend this result to the finite ring case $\G = \Zpr$, where $r$
is a positive integer and $p$ is a prime integer.
The open problem
that we solve is also mentioned in the 2007 paper~\cite{soleS07}. We first tailor the concept of encoder to the $\Zpr$ case,
making use of the specific algebraic finite chain structure of $\Zpr$. This leads to concepts of ``$p$-encoder'' and ``minimal $p$-encoder''. We then show how to construct a minimal $p$-encoder from a polynomial encoding of the code. The minimal $p$-encoder translates immediately into a minimal trellis realization. Thus our results allow for easy construction of a
minimal trellis representation from a polynomial encoding and parallel the field case.
\ruimte Convolutional codes over rings were
introduced in~\cite{masseyM89,masseyM90} where they are motivated for use
with phase modulation. In particular, convolutional codes over the
ring $\Z_M$ are useful for $M$-ary phase modulation (with $M$ a
positive integer). By the Chinese Remainder Theorem, results
on codes over $\Zpr$ can be extended to codes over $\Z_M$,
see also \cite{mcdonald74,artin91,chenCL94,johannessonWW98}.
\ruimte
Most of the literature on convolutional codes over rings adopts an
approach in which code sequences are semi-infinite Laurent
series~\cite{forneyT93,mittelholzer93,loeligerFMT94,loeligerM96,johannessonWW98,fagnaniZ96,wittenmarkisit98,
wittenmark98}. In order to make a connection with this literature,
we adopt this approach in our definition of a convolutional
code: a linear convolutional code $\C$ of length $n$ over $\Zpr$ is
defined as a subset of ${(\Zpr^n)}^\Z $ for which there exists a
polynomial matrix $G(z) \in \Zpr^{k\times n} [z]$, such that
\[
\C = \{ \cbold \in {(\Zpr^n)}^\Z
\st \exists \; \ubold \in {(\Zpr^k)}^\Z : \cbold = \ubold G(z) \AND
\]
\beq
\SUPP \ubold \subset [N, \infty) \;\mbox{for some integer}\; N \}.\label{eqdefC}
\eeq
Here $\SUPP \ubold$ denotes the support of
$\ubold$, i.e., the set of time-instants $t \in \Z$ for which
$u(t)$ is nonzero. Further, $z$ denotes the right shift operator
$zu(t) = u(t-1)$. Clearly,~(\ref{eqdefC})
implies that $\C$ is linear and shift-invariant with respect to
both $z$ and $z^{-1}$. If the matrix $G(z)$ has full row rank then $G(z)$ is called an {\em encoder} of $\C$.
\ruimte
For the field case any linear convolutional code admits a left prime
polynomial encoder, i.e., an encoder that has a polynomial right
inverse.
Such an encoder $G(z)$ gives rise to the following two properties:
\begin{enumerate}
\item {\em delay-free property}: for any $N \in \Z$
\[
\SUPP \cbold \subset [N, \infty) \implies \SUPP \ubold \subset [N, \infty)
\]
\item {\em noncatastrophic property}:
\[
 \SUPP \cbold \;\mbox{is finite} \implies \SUPP \ubold \;\mbox{is finite},
\]
\end{enumerate}
where $\cbold=\ubold G(z)$. Clearly, in the field case,
``delay-free''-ness and ``catastrophicity'' are encoder properties,
not code properties.
For the ring case, however, there are codes that do not admit a
noncatastrophic encoder. For example
(see~\cite{forneyT93,mittelholzer93,fagnaniZ01}) the convolutional
code over $\Z_4$ with encoder $G(z) = \twee{1+z}{1+3z}$ does not
admit a noncatastrophic encoder.
Similarly, the rotationally invariant convolutional
code over $\Z_4$ with encoder $G(z) = \twee{3+3z+3z^2}{3+z+z^2}$ does not admit a noncatastrophic encoder. The reader is referred to~\cite{masseyM90} for motivation and characterization of rotationally invariant codes over rings.
Further, there are codes that do not admit a delay-free encoder. For example
(see~\cite{masseyM90,loeligerM96,fagnaniZ01}) the convolutional code over $\Z_4$ with encoder $G(z) =
\twee{2}{2+z}$ does not admit a delay-free encoder. Note
that some codes over $\Zpr$ do not even admit an encoder, for
example over $\Z_4$ the code given by (\ref{eqdefC}) with
\[
G(z) = \bmat{ccc} 1+z & z & z^2 \\ 2 & 2 & 2 \emat .
\]
The literature (see e.g. \cite[subsect. V-C]{fagnaniZ01}) has declared the properties of
``delay-free'' and ``catastrophic'' to be properties of the code
rather than the encoding procedure.
By resorting to a particular type of polynomial encoder, named ``$p$-encoder'', we show in
section \ref{sec_main} that delay-freeness is not a property of the
code but of the encoding procedure, just as in the field case,
see also~\cite{kuijperPcontrolo08}. We conjecture
that the same is true for catastrophicity. To support this argument,
in section \ref{sec_con} we examine specific catastrophic convolutional
codes over $\Zpr$ and show that a noncatastrophic $p$-encoder
exists for these examples.
\ruimte
A more recent approach~\cite{rosenthalSY96} (see
also~\cite{gluesingS07, soleS07}) to convolutional codes focuses on
so-called ``finite support convolutional codes'' in which the input
sequence $\ubold$ corresponds to a polynomial. Thus the natural time
axis is $\Z_+$ and both input sequences and code sequences have
finite support. Finite support convolutional codes are, by
definition, noncatastrophic (Property 2 above) and can be
interpreted as submodules of $\mathbb Z_{p^r}^n[z]$. For $n=1$
connections can be made with polynomial block codes.
For more details the reader is
referred to our paper~\cite{kuijperPcastle08}.
\section{Preliminaries}\label{sec_prelim}
A set that plays a fundamental role
throughout the paper is the set of ``digits", denoted by
$\A_p = \{0,1,\dots, p-1\} \subset \mathbb Z_{p^r}$.
Recall that any element $a \in \dubbel{Z}_{p^r}$ can be written uniquely as
$a=\theta_0 +\theta_1 p + \cdots +  \theta_{r-1} p^{r-1}$, where $\theta_\ell \in
\A_p$ for $\ell =0 , \ldots , r-1$ ($p$-adic expansion).
This fundamental property of the ring $\Zpr$ essentially expresses a
type of linear independence among the elements $1$, $p$, $p^2$, ..., $p^{r-1}$.
It leads to specific notions of ``$p$-linear independence'' and
``$p$-generator sequence'' for
modules in $\Zpr^n$, as developed in the 1996 paper~\cite{vaziraniSR96}. For example,
for the simplest case $n=1$, the elements $1$, $p$, $p^2$, ..., $p^{r-1}$ are
called ``$p$-linearly independent'' in~\cite{vaziraniSR96} and the module $\Zpr =
\SPAN \{ 1 \}$ is written as $\Zpr = \PSPAN \{ 1 , p , p^2 , \dots
, p^{r-1} \}$. The module $\Zpr$ is said to have ``$p$-dimension'' $r$.
\ruimte
In this section we recall the main concepts
from~\cite{kuijperPP07} on modules in $\mathbb Z_{p^r}^n[z]$, that
are needed in the sequel. We present the notions of $p$-basis
and $p$-dimension of a submodule of $\mathbb Z_{p^r}^n[z]$, which
are extensions from~\cite{vaziraniSR96}'s notions for submodules
of $\mathbb Z_{p^r}^n$.  From~\cite{kuijperPP07} we also recall the concept of a reduced $p$-basis in
$\mathbb Z_{p^r}^n[z]$ that plays a crucial role in the next section.
\ruimte
\begin{definition} ~\cite{kuijperPP07}
Let $\{v_1(z), \dots, v_m(z) \} \subset \mathbb Z_{p^r}^n[z]$. A
{\bf $\boldsymbol{p}$-linear combination} of $v_1(z), \dots,
v_m(z)$ is a vector $\displaystyle \sum_{j=1}^m a_j(z) v_j(z),$
where $a_j(z) \in \mathbb Z_{p^r}[z]$ is a polynomial with
coefficients in $\A_p$ for $j=1, \dots, m$. Furthermore, the set of
all $p$-linear combinations of $v_1(z), \dots, v_m(z)$ is denoted by {\bf $\boldsymbol{p}$-span}$(v_1(z), \dots, v_m(z))$, whereas the set of all linear
combinations of $v_1(z), \dots, v_m(z)$ with coefficients in
$\mathbb Z_{p^r}[z]$ is denoted by $\SPAN (v_1(z), \dots, v_m(z))$.
\end{definition}
\ruimte
\begin{definition} ~\cite{kuijperPP07}
A sequence $(v_1(z), \dots, v_m(z))$ of vectors in $\mathbb Z^n_{p^r}[z]$ is said to be a
{\bf $\boldsymbol{p}$-generator sequence} if $p \, v_m(z)=0$ and $p \, v_i(z)$ is a $p$-linear combination of $v_{i+1}(z),
\dots, v_m(z)$ for $i=1, \dots, m-1$.
\end{definition}
\ruimte
The next lemma is a straightforward result that is used in
section~\ref{sec_main}.
\begin{Lemma}\label{lemma_atzero}
Let $(v_1(z), \dots, v_m(z) )$ be a $p$-generator sequence in $\mathbb
Z_{p^r}^n[z]$. Then $(v_1(0), \dots, v_m(0) )$ is a $p$-generator sequence in $\mathbb
Z_{p^r}^n$.
\end{Lemma}
\ruimte
\begin{Stel} ~\cite{kuijperPP07} Let $v_1(z), \dots, v_m(z) \in \mathbb Z^n_{p^r}[z]$. If
$(v_1(z), \dots, v_m(z))$ is a $p$-generator sequence then
\[
\PSPAN (v_1(z), \dots, v_m(z))= \SPAN (v_1(z), \dots, v_m(z)) .
\]
In particular, $\PSPAN (v_1(z), \dots, v_m(z))$ is a submodule
of $\mathbb Z^n_{p^r}[z]$.
\end{Stel}
\ruimte
\begin{definition} ~\cite{kuijperPP07}
The vectors $v_1(z), \dots, v_m(z) \in \mathbb Z_{p^r}^n[z]$ are
said to be {\bf $\boldsymbol{p}$-linearly independent} if the only
$p$-linear combination of $v_1(z), \dots, v_m(z)$ that equals zero is the trivial one.
\end{definition}
\ruimte
\begin{definition}
Let $M$ be a submodule of $\dubbel{Z}_{p^r}^n [z ]$, written as a
  $p$-span  of a $p$-generator sequence $(v_1(z) , v_2(z) , \cdots
  , v_m(z) )$. Then  $(v_1(z) , v_2(z) , \cdots , v_m(z) )$ is
  called a
  {\bf $\boldsymbol{p}$-basis} for $M$ if the vectors $v_1(z), \dots,
  v_m(z)$ are $p$-linearly independent in $\dubbel{Z}_{p^r}^n[z]$.
\end{definition}
\ruimte
\begin{Lemma}\label{lemma_unique}~\cite{kuijperPP07}
Let $M$ be a submodule of $\dubbel{Z}_{p^r}^n [z ]$ and let
$(v_1(z) ,
  v_2(z) , \cdots   , v_m(z) )$ be a $p$-basis for $M$. Then each vector of
$M$ is written in a unique way as a $p$-linear combination of
$v_1(z), \dots, v_m(z)$.
\end{Lemma}
\ruimte
All submodules of $\mathbb Z_{p^r}^n[z]$ can be written as the
$p$-span of a $p$-generator sequence. In fact, if
$M= \SPAN (g_1(z), \dots, g_k(z))$ then $M$ is the $p$-span of the
$p$-generator sequence
\[
(g_1(z),p g_1(z), \dots, p^{r-1}g_1(z), \dots,
g_k(z),  \dots, p^{r-1}g_k(z)) .
\]
Next, we recall a particular $p$-basis for a submodule of $\mathbb
Z_{p^r}^n[z]$, called ``reduced $p$-basis''. We first recall the
concept of ``degree'' of a vector in $\mathbb Z_{p^r}^n[z]$,
which is the same as in the field case.
\ruimte
\begin{definition}
Let $v(z)$ be a nonzero vector in $\mathbb Z_{p^r}^n[z]$, written
as $v(z)=v_0 + v_1 z + \cdots + v_d z^d$, with $v_i \in \mathbb
Z_{p^r}^n$, $i=0, \dots, d$, and $v_d \neq 0$. Then $v(z)$ is said
to have {\bf degree} $d$, denoted by $\DEG v(z)=d$.
Furthermore, $v_d$ is called the {\bf leading coefficient vector} of $v(z)$, denoted by $v^{lc}$.
\end{definition}
\ruimte
In the sequel, we denote the {\em leading row coefficient
matrix} of a polynomial matrix $V(z)$ by $V^{lrc}$. A matrix $V(z)$ is called {\bf row-reduced} if $V^{lrc}$ has full row rank.
\ruimte
\begin{Lemma}~\cite{kuijperPP07}
Let $M$ be a submodule of $\dubbel{Z}_{p^r}^n [z ]$, written as a
$p$-span of a $p$-generator sequence $(v_1(z), \dots, v_m(z))$
with $v_1^{lc}, \dots, v_m^{lc}$ $p$-linearly independent in $\mathbb Z_{p^r}^n$.
Then $(v_1(z), \dots, v_m(z))$ is a $p$-basis for $M$.
\end{Lemma}
\ruimte
\begin{definition} ~\cite{kuijperPP07}
Let $M$ be a submodule of $\dubbel{Z}_{p^r}^n [z ]$, written as a
$p$-span of a $p$-generator sequence $(v_1(z), \dots, v_m(z))$.
Then $(v_1(z),
  \dots, v_m(z))$ is called a {\bf reduced $\boldsymbol{p}$-basis} for
  $M$ if the vectors $v_1^{lc}, \dots, v_m^{lc}$ are
$p$-linearly independent in $\mathbb Z_{p^r}^n$.
\end{definition}
\ruimte
A reduced $p$-basis in $\mathbb Z^n_{p^r}[z]$ generalizes the
concept of row reduced basis from the field case. Moreover, it also
leads to the predictable degree property and gives
rise to several invariants of $M$, see~\cite{kuijperPP07}. In particular, the number of
vectors in a reduced $p$-basis as well as the degrees of these
vectors (called {\bf $\boldsymbol{p}$-degrees}), are invariants of $M$. Consequently, their sum is
also an invariant of $M$.
\ruimte
Every submodule $M$ of $\mathbb Z_{p^r}^n[z]$ has a reduced $p$-basis.
A constructive proof is given by Algorithm 3.11
in~\cite{kuijperPP07} that takes as its input a set of spanning vectors
and produces a reduced $p$-basis of $M$.
It is easy to see that if the input
is already a $p$-basis of $m$ vectors, then
the algorithm produces a reduced $p$-basis of again $m$ vectors.
Since $m$ is an invariant of the module, it follows that all
$p$-bases of $M$ have the same number of elements. As a result, the
next definition is well-defined and not in conflict with the
slightly different definition of~\cite{kuijperPP07}. \ruimte
\begin{definition} The number of elements of a $p$-basis of a submodule
$M$ of $\mathbb Z_{p^r}^n[z]$ is called the {\bf $\boldsymbol{p}$-dimension} of $M$, denoted as $\PDIM (M)$.
\end{definition}
\noindent
In recent work~\cite{kuijperS09} it is shown that computational
packages for computing minimal Gr\"obner bases can be used to
construct a minimal $p$-encoder.
\section{Minimal trellis construction from a $p$-encoder}\label{sec_main}
Formally, we define a {\em trellis section} as a
three-tuple $X = (\Zpr^n , S , K)$, where $S$ is the {\em trellis state
  set} and $K$ is the {\em set of
  branches} which is a subset of $S \times \Zpr^n \times S$, see
also~\cite{forneyT93,loeligerM96}.
A {\em trellis} is a sequence $\X = \{ X_t \}_{t\in
{\mathbb Z}}$ of trellis sections $X_t = ( \Zpr^n , S , K_t)$. A {\em
path} through the trellis is a sequence $( \cdots , b_{t-1} , b_t ,
b_{t+1} , \cdots )$ of branches $b_t = (s_t, c_t , s_{t+1}) \in K_t$
such that $b_{t+1}$
starts in the trellis state where $b_t$ ends for $t \in \Z$.
The set of all trellis paths that start at the
zero state is denoted by $\pi (\X)$.
The mapping $\lambda : \pi (\X) \mapsto {(\Zpr^n)}^\Z$ assigns to
every path $( \cdots , b_{t-1} , b_t ,b_{t+1} , \cdots )$ its label
sequence $( \cdots , c_{t-1} , c_t ,c_{t+1} , \cdots )$. A trellis $\X$ is called a
{\em trellis representation} for a convolutional code $\C$ if $\C = \lambda (\pi (\X))$.
\ruimte
A trellis representation $\X$ for a convolutional code $\C$ is called
{\em minimal} if the size of its trellis state set $S$ is minimal
among all trellis representations of $\C$. It is wellknown how to
construct a minimal trellis representation in terms of the code
sequences of $\C$. In fact, the theory of canonical trellis
representations from the field case carries through to the ring case,
see~\cite{wil88,forneyT93,loeligerM96}. Since it plays a crucial role
in the proof of our main result, we recall the definition of canonical
trellis in Appendix A.
\ruimte
Let us
recall the wellknown controller canonical form. Let $\R$ be a ring.
A matrix $E(z) \in \R^{\kappa \times n} [z]$ is realized in
controller canonical form~\cite{kailath} (see also~\cite[Sect.
5]{fornasiniP04}) as \beq E(z)= B {(z^{-1}I - A)}^{-1}C + D ,
\label{eq_concanform} \eeq as follows. Denoting the $i$'th row of
$E(z)$ by $e_i(z) = \sum _{\ell=0}^{\delta_i} e_{i,\ell}z^\ell$,
where $e_{i,\ell} \in \R^{1\times n}$
and $e_{i,\delta_i} \neq 0$,
the matrices $A$, $B$, $C$ and $D$ in
(\ref{eq_concanform}) are given by
\[
A = \bmat{ccc}A_1 & & \\ & \ddots & \\ && A_{\kappa} \emat,
\;\;\;\; B = \bmat{ccc}B_1 && \\ & \ddots & \\ && B_{\kappa}
\emat,
\]
\[
\;\;\;\; C = \bmat{c} C_1
\\ \vdots \\ C_{\kappa} \emat, \;\;\;\; D = \bmat{c} e_{1,0}
\\ \vdots \\ e_{\kappa,0} \emat ,
\]
where
$A_i$ is a $\delta_i \times \delta_i$ matrix, $B_i$ is a $1 \times \delta_i$ matrix and
$C_i$ is a $\delta_i \times 1$ matrix, given by
 \[ A_i = \bmat{cccc}    0    &   1    &        &    \\
                                    & \ddots & \ddots &    \\
                                    &        & \ddots &  1  \\
                                    &        &        &  0 \emat, \;\;\;\;
                                    B_i = \bmat{cccc} 1 & 0 & \cdots &
                                    0 \emat,
                                    \]
                                    \beq \;\;\;\; C_i = \bmat{c}
                                    e_{i,1}\\ \vdots \\
                                    e_{i,\delta_i}\emat \;\;
                                    \mbox{for} \;\; i=1, \ldots ,
                                    \kappa.\label{canABCD}
\eeq Whenever $\delta_i = 0$, the $i$th block in $A$ as well as
$C$ is absent and a zero row occurs in $B$. Denoting the sum of
the $\delta_i$'s by $\delta$, it is clear that $A$ is a $\delta
\times \delta$ nilpotent matrix. The above controller canonical
realization can be visualized as a feedforward shift-register with $\delta$
registers.
\ruimte In the case that $\R$ is a field with $q$ elements it is
wellknown~\cite{johannessonW93,loeligerM96} how to obtain a minimal
trellis representation for $\C$ from a polynomial encoder. For this,
the rows of the polynomial encoder should first be algebraically
manipulated (using Smith form and row reduction operations) to yield
a left prime and row reduced encoder $G(z)$. Then $G(z)$ is called
{\em canonical} in the literature, see~\cite[App.\ II]{loeligerM96}.
A minimal trellis representation of $\C$ is then provided by the
controller canonical realization $G(z) = B {(z^{-1}I - A)}^{-1}C +
D$ as in~(\ref{canABCD}).
Although this result is known, in Appendix B we give a proof by
showing that there exists an isomorphism between the trellis state
set of the controller canonical realization and the trellis state
set of the canonical trellis (as defined in Appendix A) of $\C$. The
set is thus minimal and has $q^\nu$ elements, where $q$ is the
number of elements of the field and $\nu$ is the sum of the row
degrees of $G(z)$. The invariant $\nu$ is commonly referred to as
the ``degree'' of the code $\C$ (but called the ``overall constraint
length'' in the early literature). The row degrees are called the
``Forney indices'' of the code~\cite{mceliece98}. \ruimte
Below we consider convolutional codes over $\Zpr$ that admit
a noncatastrophic encoder, for simplicity, we call such codes
noncatastrophic. We show that such codes admit a particular
type of polynomial encoder (later called ``minimal
$p$-encoder''), whose controller canonical realization provides a
minimal trellis representation, just as in the field case. We are
then also able to express the minimal number of trellis states in
terms of the sum of the row degrees of a minimal $p$-encoder.
\ruimte Let us now first introduce the notion of ``$p$-encoder''.
Recall that $\A_p = \{ 0 , 1 , \ldots , p-1 \} \subset \Zpr$.
\ruimte
\begin{definition}\label{def_p-encoder}
Let $\C$ be a convolutional code of length $n$ over
$\Zpr$. Let $E(z) \in \Zpr^{\kappa\times n} [z]$ be a polynomial matrix
whose rows are a $p$-linearly independent $p$-generator sequence. Then
$E(z)$ is said to be a
{\bf $\boldsymbol{p}$-encoder} for $\C$ if
\[
\C = \{ \cbold \in {(\Zpr^n)}^\Z \st \exists \; \ubold \in
{(\A_p^{\kappa})}^\Z : \cbold = \ubold E(z) \AND
\]
\[
\;\;\;\;\;\;\SUPP \ubold \subset [N, \infty) \;\mbox{for some
    integer}\; N \} .
\]
The integer $\kappa$ is called the {\bf $\boldsymbol{p}$-dimension} of
$\C$. Furthermore, $E(z)$ is said to be a {\bf delay-free} $p$-encoder
if for any $N \in \Z$ and any $\cbold \in \C$, written as $\cbold =
\ubold E(z)$ with $\ubold \in {(\A_p^{\kappa})}^\Z$ we have
\[
\SUPP \cbold \subset [N, \infty) \implies \SUPP \ubold \subset [N, \infty) .
\]
Also, $E(z)$ is said to be a {\bf noncatastrophic} $p$-encoder if for any $\cbold \in \C$, written as $\cbold = \ubold E(z)$ with $\ubold \in {(\A_p^{\kappa})}^\Z$ we have
\[
 \SUPP \cbold \;\mbox{is finite} \implies \SUPP \ubold \;\mbox{is finite}.
\]
Finally, a convolutional code $\C$ that admits a noncatastrophic $p$-encoder is called {\bf noncatastrophic}.
\end{definition}
\ruimte
Thus a difference between a $p$-encoder $E(z)$ and the encoding matrix $G(z)$ of~(\ref{eqdefC}), is that the inputs of $E(z)$ take their values in $\A_p$ rather than in
$\Zpr$. Note that the idea of using a $p$-adic expansion for the
input sequence is already present in the 1993
paper~\cite{forneyT93}. It was not until 1996 that the crucial notion
of $p$-generator sequence appeared in~\cite{vaziraniSR96}, but only
for constant vectors --- it was extended to polynomial vectors
in~\cite{kuijperPP07}. In our
definition the rows of a $p$-encoder are required to be a
$p$-generator sequence consisting of polynomial vectors.
\ruimte
Recall that a convolutional code over $\Zpr$ is given by~(\ref{eqdefC}):
\[
\C = \{ \cbold \in {(\Zpr^n)}^\Z
\st \exists \; \ubold \in {(\Zpr^k)}^\Z : \cbold = \ubold G(z) \AND
\]
\[
\SUPP \ubold \subset [N, \infty) \;\mbox{for some integer}\; N \}.
\]
Also recall that there exist convolutional codes over $\Zpr$
that do not admit a $G(z)$ of full row rank, i.e. an encoder. An
important observation is that {\em any} convolutional code over $\Zpr$ admits a $p$-encoder, even a $p$-encoder $E(z)$, such that the rows of $E^{lrc}$ are $p$-linearly independent in
$\Zpr^n$.
Indeed, any reduced $p$-basis of the polynomial module spanned by
the rows of $G(z)$, produces the rows of such a
$p$-encoder $E(z)$. This shows that the concept of $p$-encoder is
more natural than the concept of encoder as it is tailored to the
algebraic structure of $\Zpr$. The next lemma is straightforward.
\ruimte
\begin{Lemma}
Let $E(z) \in \Zpr^{\kappa\times n} [z]$ be a $p$-encoder for a
convolutional code $\C$ of length $n$. Then $E(z)$ is delay-free
property (Definition~\ref{def_p-encoder}) if and only if the
rows of $E(0)$ are $p$-linearly independent in $\Zpr^n$.
\end{Lemma}
\ruimte
\begin{Stel} \label{delay-free}
Let $\C$ be a convolutional code of length $n$ over $\Zpr$. Then
$\C$ admits a delay-free $p$-encoder $E(z) \in \Zpr^{\kappa\times n}
[z]$ for some integer $\kappa$, such that the rows of $E^{lrc}$ are $p$-linearly independent in
$\Zpr^n$.
\end{Stel}
\begin{IEEEproof}
As noted above, $\C$ admits a $p$-encoder $E(z)$, such that the rows
of $E^{lrc}$ are $p$-linearly independent in $\Zpr^n$, i.e., they
constitute a reduced $p$-basis. Without loss of generality we may
assume that the row degrees of $E(z)$ are nonincreasing. Let $L$ be the smallest nonnegative integer such that the
last $\kappa -L$ rows of $E(z)$ are a delay-free $p$-encoder.
\ruimte Now assume
that $L >0$ (otherwise we are done). If $L=\kappa$ it means that the
last row $e_\kappa (z)$ of $E(z)$
can be written as
\[
e_\kappa (z) = z^\ell \bar e_\kappa (z),
\]
where $\ell > 0$ and $\bar e_\kappa (z) \in \Zpr^n[z]$ with $\bar
e_\kappa (0)  \neq 0$. Note that $\deg \bar e_\kappa (z) < \deg
e_\kappa (z)$.
Clearly, $(e_1(z) , \ldots , e_{\kappa -1}(z), \bar e_\kappa (z))$
is
a $p$-encoder of $\C$, whose rows are still a reduced $p$-basis.

\ruimte

If $L < \kappa$, then, by construction, there exist $\alpha_j \in
{\cal A}_p$ for
$j=L+1, \dots, \kappa$, such that
$$e_L (0) + \sum_{j>L} \alpha_j e_j (0) = 0$$
(use the fact that $(e_1 (0), \ldots , e_\kappa (0))$ is a
$p$-generator sequence by Lemma~\ref{lemma_atzero}). Replacing
$e_L(z)$ by $\tilde e_L (z) := e_L (z) + \sum_{j>L} \alpha_j e_j
(z)$ obviously gives a $p$-basis $(e_1(z) , \ldots ,e_{L-1} (z),
\tilde e_L (z), e_{L+1} (z) , \ldots , e_\kappa (z))$ of the module
spanned by $e_1(z) , \ldots , e_L (z) , \ldots , e_\kappa (z)$
and, consequently, a $p$-encoder of $\C$. Moreover, by the
$p$-predictable degree property (Theorem 3.8 of~\cite{kuijperPP07}),
$\deg \tilde e_L (z) = \deg e_L (z)$,
which means that
$(e_1(z), \dots, \tilde e_L(z), \dots, e_\kappa (z))$ is still a
reduced $p$-basis. Since $\tilde e_L (0) = 0$, we can write $\tilde
e_L (z) = z^{\tilde \ell} \bar
e_L (z)$, with $\bar e_L (0) \neq 0$ and $\tilde \ell > 0$. Note
that $p\tilde e_L (z)$ is a $p$-linear combination $p\tilde e_L (z)
= \sum_{j>L} \beta_j (z) e_j (z)$ with $\beta_j (z) \in \A_p [z]$.
Because of the $p$-linear independence of $e_{L+1} (0) , \ldots ,
e_\kappa (0)$, we must have that the coefficients $\beta_j(z)$ are
of the form $\beta_j(z) = z^{\ell_j} \bar \beta_j(z)$ with $\ell_j
\geq \tilde \ell$ for $L+1 \leq j \leq \kappa$. Consequently, the
sequence $(e_1(z) , \ldots, e_{L-1} (z) ,\bar e_L (z),e_{L+1} (z) ,
\ldots , e_\kappa (z))$ is a $p$-encoder of $\C$, which is still a
reduced $p$-basis with $\deg \bar e_L (z) < \deg e_L (z)$. If
$(e_1(z) , \ldots ,\bar e_L (z),e_{L+1} (z) ,\ldots , e_\kappa (z))$
is not a delay-free $p$-encoder,
then re-order the vectors so that their degrees are nonincreasing
and repeat this procedure until a
delay-free $p$-encoder for
$\C$ is obtained.
Since the sum of the row degrees of $p$-bases obtained at each step
of the procedure is lower than in the previous step, a delay-free
$p$-encoder is obtained after finitely many iterations.
\end{IEEEproof}
The next example is a simple example that illustrates the above theorem.
\begin{Voorb}\label{ex_delay}
Over ${\mathbb Z}_4$: consider the $(2,1)$ convolutional code $\C$ of~\cite[p. 1668]{loeligerM96} given by
the polynomial encoder
\[
G(z) = \twee{2}{2+z} .
\]
A delay-free $p$-encoder for $\C$ is given by
\[
E(z) = \bmat{cc}2 & 2+z \\ 0 & 2 \emat .
\]
\end{Voorb}
\begin{Stel} \label{del-and-noncat}
Let $\C$ be a noncatastrophic convolutional code of length $n$ over $\Zpr$. Then
$\C$ admits a delay-free noncatastrophic $p$-encoder $E(z) \in \Zpr^{\kappa\times n}
[z]$ for some integer $\kappa$, such that the rows of $E^{lrc}$ are $p$-linearly independent in
$\Zpr^n$.
\end{Stel}
\begin{IEEEproof}
By definition there exists a noncatastrophic $p$-encoder $E_1(z)$
for $\C$. Apply Algorithm 3.11 of~\cite{kuijperPP07} to the rows of
$E_1(z)$. This gives us a reduced $p$-basis $e_1 (z) , \ldots ,
e_\kappa (z)$ for the module spanned by the rows of $E_1(z)$. Define
$E_2(z)$ as the $\kappa\times n$ polynomial matrix with $e_1 (z) ,
\ldots , e_\kappa (z)$ as rows. By construction the rows of
$E_2^{lrc}$ are $p$-linearly independent in $\Zpr^n$. It is easy to
see that $E_2(z)$ is still noncatastrophic.
If $E_2(z)$ is not delay-free apply the procedure of the proof of
Theorem \ref{delay-free} to $E_2(z)$ to obtain a delay-free
$p$-encoder $E(z)$, such that the rows of $E^{lrc}$ are $p$-linearly
independent in $\Zpr^n$. It is easy to see that $E(z)$ is still
noncatastrophic.
\end{IEEEproof}
\begin{definition}
Let $\C$ be a noncatastrophic convolutional code of length $n$ over $\Zpr$.
Let $E(z) \in \Zpr^{\kappa\times n} [z]$ be a delay-free noncatastrophic $p$-encoder
for $\C$, such
that the rows of $E^{lrc}$ are $p$-linearly independent in
$\Zpr^n$. Then $E(z)$ is called a {\bf minimal
  $\boldsymbol{p}$-encoder} of $\C$.
Furthermore, the {\bf $\boldsymbol{p}$-indices} of $\C$ are defined as
the row degrees of $E(z)$ and the {\bf $\boldsymbol{p}$-degree} of $\C$ is
defined as the sum of the $p$-indices of $\C$.
\end{definition}
\ruimte
Thus, in the terminology of section~\ref{sec_prelim}, the rows of a
minimal $p$-encoder are a reduced $p$-basis. If the code
$\C$ has a canonical encoder $G(z)$, then both $G^{lrc}$ mod $p$ and
$G(0)$ mod $p$ have full row rank in $\Z_p^{k\times n}$, so that
a minimal $p$-encoder is
trivially constructed as \beq E(z) = \bmat{c} G(z) \\ pG(z) \\
\vdots \\ p^{r-1} G(z) \emat.\label{p-enc_trivial} \eeq
An important
observation is that all noncatastrophic codes admit a minimal
$p$-encoder $E(z)$ but not all
such codes admit an encoder $G(z)$ that is row reduced and/or delay-free.
\begin{definition}\label{def_ABCDcan_ring}
Let $\C$ be a convolutional code of length $n$ with $p$-encoder $E(z) \in
\Zpr^{\kappa\times n} [z]$. Denote the sum of the row degrees of $E(z)$ by
$\gamma$ and let
\[
(A,B,C,D) \in \Zpr^{\gamma  \times \gamma} \times \Zpr^{\kappa
  \times \gamma}\times \Zpr^{\gamma  \times n} \times \Zpr^{\kappa  \times
  n}
\]
be a controller canonical realization of $E(z)$.
Then the {\bf controller canonical trellis} corresponding to $E(z)$ is
defined as $\X = \{ X_t \}_{t\in {\mathbb Z}}$, where $X_t = (\Zpr^n ,
\A_p^\gamma,K_t)$ with
\[
K_t = \{ (s(t) , s(t) C + u(t)D , s(t)A + u(t)B \;\mbox{ such that}
\]
\[
\;\;\;\;\;  s(t) \in \A_p^\gamma , \;\;\; u(t) \in \A_p^\kappa \} .
\]
\end{definition}
Note that the states take their values in the nonlinear set
$\A_p^{\gamma}$, which is not closed with respect to addition or
scalar multiplication. Similarly, the inputs take their values in the
nonlinear set $\A_p^{\kappa}$.
The next theorem presents our main result.
\ruimte
\begin{Stel}\label{thm_min}
Let $\C$ be a noncatastrophic convolutional code of length $n$ with minimal $p$-encoder $E(z) \in
\Zpr^{\kappa\times n} [z]$. Denote the $p$-degree of $\C$ by $\gamma$.
Then the controller canonical trellis corresponding to $E(z)$ is a minimal trellis
representation for $\C$. In particular, the minimum number of
trellis states equals $p^\gamma$.
\end{Stel}
\begin{IEEEproof}
see Appendix B.
\end{IEEEproof}
\ruimte In the field case $r=1$ the above theorem coincides with the
classical result, i.e., the minimum number of trellis states equals
$p^{\gamma}$, where $\gamma$ is the degree of the code. \ruimte
For convolutional codes that admit a canonical encoder, we have the
following corollary, which follows immediately from applying Theorem~\ref{thm_min} to the minimal
$p$-encoder given by~(\ref{p-enc_trivial}).
Note that the result coincides with results in~\cite[Sect.
7.4]{wittenmark98}, where a canonical encoder is called
``minimal-basic''.
\begin{Gevolg}
Let $\C$ be a $(n,k)$ convolutional code that has a canonical
encoder $G(z) \in \Zpr^{k\times n} [z]$. Then the $rk$ $p$-indices
of $\C$ are the $k$ row degrees of $G(z)$, each occurring $r$ times.
The minimum number of trellis states equals $q^\nu$, where $\nu$ is
the sum of the row degrees of $G(z)$ and where $q=p^r$.
\end{Gevolg}
The next example illustrates our theory for the more interesting case
where the code does not admit a canonical encoder.
\ruimte
\begin{Voorb}
Over ${\mathbb Z}_4$: consider the $(3,2)$ convolutional code $\C$ given by
the polynomial encoder
\[
G(z) = \bmat{c}g_1(z) \\ g_2(z) \emat , \;\;\;\mbox{where}
\]
\[
\;\; g_1(z)
= \bmat{ccc} z^2 + 1  & 1 & 0 \emat \AND g_2(z) = \bmat{ccc} 2z & 2 & 1 \emat .
\]
Clearly, $G(z)$ is a left prime encoder whose controller canonical
trellis has $4^3=64$ trellis states. Note that $G^{lrc}$ does not
have full row rank and therefore $G(z)$ is not canonical. Denote by im
$G(z)$ the polynomial module spanned by the  
rows of $G(z)$. A
$p$-basis for the module im $G(z)$ is provided by the rows of the
matrix
\[
\bmat{c} g_1(z) \\ 2 g_1(z) \\ g_2(z) \\ 2 g_2(z) \emat = \bmat{ccc}
z^2 + 1  & 1 & 0 \\ 2z^2 + 2  & 2 & 0  \\ 2z & 2 & 1 \\ 0 & 0 & 2
\emat ,
\]
which has leading row coefficient matrix
\[
\bmat{ccc}
1  & 0 & 0 \\ 2  & 0 & 0  \\ 2 & 0 & 0\\ 0 & 0 & 2 \emat .
\]
The row reduction algorithm of~\cite[Algorithm 3.11]{kuijperPP07}
is particularly simple in this case: by adding $z$ times the third row to the second row, we obtain the
matrix $E(z)$, given by
\[
E(z) = \bmat{ccc}
z^2 + 1  & 1 & 0 \\ 2  & 2z+2 & z  \\ 2z & 2 & 1 \\ 0 & 0 & 2
\emat ,
\]
whose rows are a reduced $p$-basis for the module im $G(z)$. Indeed,
the rows of its leading row coefficient matrix, given by
\[
\bmat{ccc}
1  & 0 & 0 \\ 0  & 2 & 1  \\ 2 & 0 & 0\\ 0 & 0 & 2 \emat ,
\]
are $p$-linearly independent. As a result, the $p$-indices of $\C$ are
$2$, $1$, $1$, $0$ and the $p$-degree of $\C$ equals $4$.
The controller canonical trellis corresponding to $E(z)$ is given by
\[
A=\bmat{cccc} 0 & 1 & 0 & 0 \\ 0 & 0 & 0 & 0 \\ 0 & 0 & 0 & 0 \\ 0 & 0
& 0 & 0 \emat ; \;\;
B=\bmat{cccc} 1 & 0 & 0 & 0\\ 0 & 0 & 1 & 0 \\ 0 & 0 & 0 & 1 \\ 0 & 0 &
0 & 0 \emat ;
\]
\[
C =\bmat{ccc} 0 & 0 & 0 \\ 1 & 0 & 0 \\ 0 & 2 & 1 \\ 2 & 0 & 0 \emat ; \;\;
D=\bmat{ccc} 1 & 1 & 0 \\ 2 & 2 & 0 \\ 0 & 2 & 1 \\ 0 & 0 & 2 \emat .
\]
This trellis is minimal with $2^4 = 16$ trellis states.
\end{Voorb}
\begin{Voorb}
Over ${\mathbb Z}_4$: consider the $(2,1)$ convolutional code $\C$ of Example~\ref{ex_delay},
given by
the polynomial encoder $G(z) = \twee{2}{2+z}$ (note that $G(z)$ is not delay-free).
The delay-free $p$-encoder
\[
E(z) = \bmat{cc}2 & 2+z \\ 0 & 2 \emat .
\]
of Example~\ref{ex_delay} is clearly minimal, so that its corresponding trellis is minimal with $2$ states which concurs with~\cite{forneyT93}.
\end{Voorb}
\section{Conclusions}\label{sec_con}
An important class of polynomial encoders for convolutional codes over a field are the
canonical ones. Their feedforward shift register implementations are
minimal trellis representations of the code. The trellis state space
is linear. However, for convolutional codes over
the finite ring $\Zpr$, the literature has generalized this result only for
restricted cases.
In this paper we introduce the concept of {\it $p$-encoder} and
define {\it minimal $p$-encoder} for the class of noncatastrophic
convolutional codes. We show how to obtain a minimal $p$-encoder
from a polynomial encoding of the code. We show that the feedforward
shift register implementation of such a minimal $p$-encoder is a
minimal trellis representation of the code. Its trellis state space
is nonlinear. We also express the minimal number of states in terms
of the row degrees of the minimal $p$-encoder. In our view a minimal
$p$-encoder is the ring analogon of the ``canonical polynomial
encoder" from the field case. We also present the novel concepts of
$p$-indices and
$p$-degree of a code as analogons of the field notions of ``Forney
indices" and ``degree", respectively.
\ruimte
Our approach allows us to view ``delay-freeness'' as a property of the
$p$-encoder. Thus we arrive at the novel result that delay-freeness is
a property of the encoding (just as in the field case) rather than a property of the code, as
in the literature so far (see
e.g. \cite[subsect. V-C]{fagnaniZ01}). We conjecture that a similar phenomenon occurs with respect to catastrophicity, i.e., ``noncatastrophic'' is a property of the $p$-encoder, not the code. This would imply that minimal $p$-encoders can be obtained for {\em all}
convolutional codes over $\Zpr$, including the catastrophic codes. This is of particular importance for rotationally invariant catastrophic codes, see e.g. \cite{masseyM90}. It is a topic of future research to investigate this conjecture which is likely to involve a generalization of a type of ``normal form'' for polynomial matrices over $\Zpr$.
To support our conjecture, let us examine the rotationally invariant catastrophic code $\C_1$ over $\Z_4$ given by the encoder $G_1(z) = \twee{3+3z+3z^2}{3+z+z^2}$. A noncatastrophic minimal $p$-encoder for $\C_1$ is given by
\[
E_1(z) = \bmat{cc}3+3z+3z^2 & 3+z+z^2 \\ 2 & 2 \emat ,
\]
yielding a minimal trellis representation of $\C_1$ with $4$ states.
Similarly the catastrophic code $\C_2$ over $\Z_4$ with encoder $G_2(z) = \twee{1+z}{1+3z}$ has a noncatastrophic minimal $p$-encoder
\[
E_2(z) = \bmat{cc}1+z & 1+3z \\ 2 & 2 \emat ,
\]
yielding a minimal trellis representation of $\C_2$ with $2$ states.
\section{Acknowledgments}
The authors thank the reviewers for helpful comments, particularly
for alerting us to the relevance of rotationally invariant codes.
\ruimte The first author is supported in part by the Australian
Research Council; the second author is supported in part by the
Portuguese Science Foundation (FCT) through the Unidade de
Investiga\c c\~ao Matem\'atica e Aplica\c c\~oes of the University
of Aveiro, Portugal.

\appendices
\section{}
In this appendix we recall the construction of a minimal trellis for
a convolutional code $\C$ as a so-called {\em two-sided realization}
of $\C$,
see~\cite{wil88,forneyT93,mittelholzer93,loeligerFMT94,loeligerM96,wittenmark98}.
Consider two code sequences $\cbold \in \C$ and $\ctbold \in \C$.
Conform~\cite{wil88}, the {\em
  concatenation} at time $t\in \Z$ of $\cbold$ and $\ctbold$, denoted
by $\cbold \wedge_t \ctbold$, is defined as
\begin{eqnarray*}
\cbold \wedge_t \ctbold (t')  & := & \left\{
  \begin{array}{c@{\hspace{1cm}}c} c(t') & \FOR t' < t \\
\tilde c (t') & \FOR t' \geq t \end{array}\right. .
\end{eqnarray*}
The code sequences $\cbold$ and $\ctbold$ are called {\em
equivalent}, denoted by $\cbold \simeq \ctbold$, if
\[
\cbold \wedge_0 \ctbold \in C .
\]
\begin{definition}
Let $\C$ be a linear convolutional code of length $n$ over a finite
ring $\R$. The {\bf canonical trellis} of $\C$ is defined as $\X =
\{ X_t \}_{t\in {\mathbb Z}}$, where $X_t = (\R^n , S,K_t)$ with $S
:= \C\MOD \simeq$ and


\[
K_t := \{ (s(t) , c(t) , s(t+1)) \st s(t) = z^{-t}\cbold \MOD \simeq
\;\AND\;
\]
\[
\;\;\;\;\; s(t+1) = z^{-t-1}\cbold \MOD \simeq \}.
\]
\end{definition}
It has been shown in~\cite{wil88} that the above trellis is minimal.
Intuitively this is explained from the fact that, by construction,
states cannot be merged.

\section{}
In this appendix we prove Theorem~\ref{thm_min} via a bijective mapping from
the controller canonical trellis state set to the trellis state set of the
canonical trellis that is defined in Appendix A.
We first provide the
proof for the field case. In our proof of
Theorem~\ref{thm_min}, which is the ring case, we are then able to highlight the
parts that are different from the proof for the field case.
\begin{Stel}\label{thm_min_field}
Let $\C$ be a $(n,k)$ convolutional code of degree $\nu$ over a finite
field $\R$ with canonical encoder $G(z) \in
\R^{k\times n} [z]$. Then the controller canonical trellis
corresponding to $G(z)$ is a minimal trellis representation for
$\C$. In particular, the minimum number of trellis states equals
$q^\nu$, where $q$ is the size of the field $\R$.
\end{Stel}
\begin{IEEEproof}
Denote the memory of $\C$ by $\nu_*$, i.e., $\nu_*$ is the maximal
Forney index of $\C$. Consider the mapping $\Theta : \R^\nu \mapsto C\MOD
\simeq$, given by
\[
\Theta (s) := {[\cbold ]}_{\simeq} ,
\]
where $\cbold \in \C$ passes through state $s$ at time $0$. The
mapping $\Theta$ is well-defined since for any $s$ there exists such a
code sequence and any two code sequences that pass
through state $s$ at time $0$ are obviously equivalent.
\ruimte
Since the trellis state set $C\MOD \simeq$ of the canonical trellis of
Appendix A is minimal, it suffices to prove that $\Theta$ is an
isomorphism, as follows. Surjectivity follows immediately from
the fact that all code sequences pass through some state at time
$0$. Furthermore, the mapping $\Theta$ is linear since $\Theta (s_1 +
s_2) = {[\cbold_1 + \cbold_2 ]}_{\simeq}$. It remains to prove that
$\Theta$ is injective.
\ruimte
For this, let $s \in \R^\nu$ be such that $\Theta (s) = 0$. Define
$u(-\nu_*)$,...,$u(-2)$, $u(-1)$ as elements of $\R^k$ for which
\[
\bmat{cccc}u(-\nu_*) & \cdots & u(-2) & u(-1) \emat \bmat{c} B A^{\nu_*
  -1} \\ \vdots \\ BA \\ B \emat = s .
\]
Define $\ubold := (\cdots , 0 , 0 , u(-\nu_*), \cdots , u(-2) , u(-1) , 0
, 0 , \cdots )$ and let $\cbold := G(z) \ubold$ be the corresponding
code sequence. Then clearly $\cbold$ passes through $s$.
From $\Theta (s) = 0$ it now follows that the sequence
$\cbold \wedge_0 0$ is a code sequence. Denote its state at time $0$
by $s'$ and its input sequence by $\ubold'$. Then clearly
\[
\bmat{cccc}u'(-\nu_*) & \cdots & u'(-2) & u'(-1) \emat \bmat{c} B A^{\nu_*
  -1} \\ \vdots \\ BA \\ B \emat = s' .
\]
We now prove that $s=s'$, as follows. Firstly, it is clear that
\[
\bmat{cccc}c(-\nu_*) & \cdots & c(-2) & c(-1) \emat =
\]
\beq
\bmat{cccc}u(-\nu_*) &
\cdots & u(-2) & u(-1) \emat \bmat{cccc}D & BC & BAC & \cdots \\
0 & D & BC & \cdots \\
0 & 0 & D & \cdots \\
&&& \ddots \emat .\label{eq_cu}
\eeq
Furthermore, from the fact that the encoder is delay-free (Property 1 in
section~\ref{sec_intro}) it follows that $D = G(0)$ has full row rank
and that $u'(\ell) = 0$ for $\ell < -\nu_*$. As a result,
{\small \[
\bmat{cccc}c(-\nu_*) & \cdots & c(-2) & c(-1) \emat =
\]
\beq
\bmat{cccc}u'(-\nu_*) &
\cdots & u'(-2) & u'(-1) \emat \bmat{cccc}D & BC & BAC & \cdots \\
0 & D & BC & \cdots \\
0 & 0 & D & \cdots \\
&&& \ddots \emat .\label{eq_cu_prime}
\eeq}
Since $D$ has full row rank, the matrix in the above equation also has full row rank. Since
the right-hand sides of equations~(\ref{eq_cu})
and~(\ref{eq_cu_prime}) are equal, it then follows that $u(\ell) =
u'(\ell)$ for $-\nu_* \leq \ell \leq -1$. As a result $s=s'$.

\ruimte We now prove that $s=0$. By the above, $\cbold \wedge_0 0$
is a code sequence that passes through $s$ at time $0$. Its input
sequence $\ubold'$ is of the form
\[
(\cdots , 0 , 0 , u'(-\nu_*), \cdots ,  u'(M), 0 , 0 , \cdots ) ,
\]
where $M \geq 0$. Here we used the fact that the encoder is
noncatastrophic (Property 2 in section~\ref{sec_intro}). By
construction the state of $\cbold \wedge_0 0$ at time $M+\nu_*+1$
then equals zero. We now use the row reducedness of $G(z)$ to
conclude that $s=0$, as follows. Denote the state at time $M+\nu_*$
by $\bar s$. Now recall the formula~(\ref{canABCD}) for the
controller canonical form. Since $\bar s A = 0$, the nonzero
components of $\bar s$ must be last components in a $(1\times
\nu_i)$-block in $\bar s$. Also, $c(M+\nu_*)=0$, so that $\bar s C =
0$. By construction, the last rows of the $(\nu_i \times n)$-blocks
of $C$ are rows from $G^{lrc}$ and are therefore linearly
independent. As a result, $\bar s = 0$. Repeating this argument
again and again, we conclude that
$u'(0)=...=u'(M)=0$ and all states for time $\geq 0$ are zero, so
that, in particular $s =0$, which proves the theorem. Obviously, the
size of the trellis state set $S$ equals $q^\nu$.
\end{IEEEproof}

We now turn to the ring case to prove the analogon of the above
theorem. As compared to the field case, the
proof requires some care because the trellis state set $\A_p^\gamma$
is not linear.
\ruimte
\ruimte
{\bf Proof of Theorem~\ref{thm_min}:}
\ruimte
Define $\nu_*$ as the maximal $p$-index of $\C$. Consider the
mapping $\Theta : \A_p^\gamma \mapsto C\MOD \simeq$, given by
\[
\Theta (s) := {[\cbold ]}_{\simeq} ,
\]
where $\cbold \in \C$ passes through state $s$ at time $0$. Then
$\Theta$ can be shown to be well-defined and surjective, as in the proof
of Theorem~\ref{thm_min_field}.
Note that $\Theta$ is not necessarily a linear
mapping. As a result, injectivity can no longer be proven by showing
that $\Theta (s) = 0$ only for $s=0$,
as in the proof of Theorem~\ref{thm_min_field}. Thus, to show that
$\Theta$ is injective, let $s$ and $\tilde s \in \A_p^\gamma $ be
such that $\Theta (s) = \Theta (\tilde s)$. Let $\cbold$ be the code
sequence that passes through $s$ at time $0$, as defined in the
proof of Theorem~\ref{thm_min_field}. Let $\ctbold$ be the analogous
code sequence that passes through $\tilde s$ at time $0$.
Note that both $\cbold$ and $\ctbold$ have finite support. From
$\Theta (s) = \Theta (\tilde s)$ it now follows that the sequence
$\cbold \wedge_0 \ctbold$ is a code sequence. Denote its state at
time $0$ by $s'$ and its input sequence by $\ubold' \in
{(\A_p^{\kappa})}^\Z$. Since $E(z)$ is a delay-free $p$-encoder, the rows
of $E(0)$ are a $p$-basis (use also Lemma~\ref{lemma_atzero}). By Lemma 2.8
of~\cite{kuijperPP07} (see also~\cite{vaziraniSR96}), it now follows from the fact that inputs only
take values in $\A_p$ that $s=s'$. The reasoning is as in the proof of Theorem~\ref{thm_min_field}.
\ruimte
We now prove that $s=\tilde s$. By the above, $\cbold \wedge_0 \ctbold$ is a code sequence that passes
through $s$ at time $0$. As in the proof of
Theorem~\ref{thm_min_field}, it follows that its state equals zero at
time $M + \nu_* + 1$ for some $M \geq 0$. Since $E(z)$ is a
minimal $p$-encoder, the rows of $E^{lrc}$
are $p$-linearly independent. It now follows from the fact that states only
take values in $\A_p$ that the state at time $M +\nu_*$ must also be zero. The reasoning is
as in the proof of Theorem~\ref{thm_min_field}.
Repeating this argument again and again, we conclude that all states
for time $\geq \nu_*$ are zero. As a result, $u'(0) = u'(1) = \cdots =
u'(\nu_* -1) = 0$, so that
\[
s \bmat{cccc} C & AC & \cdots & A^{\nu_* -1}C \emat
\]
\[
\;\;\;\;\; = \tilde s \bmat{cccc}
C & AC & \cdots & A^{\nu_* -1}C \emat .
\]
We now prove that the above equation implies that $s=\tilde s$. By
Theorem 3.10 of~\cite{kuijperPP07}, the rows of $E^{lrc}$
are not only $p$-linearly independent but also a $p$-generator
sequence. By Lemma 2.8 of~\cite{kuijperPP07} any $p$-linear combination of these rows
is then unique. By construction, this property is inherited by the rows of $\bmat{cccc} C & AC & \cdots &
A^{\nu_* -1}C \emat$ . Since both $s$ and $\tilde
s$ take their values in $\A_p$, it therefore follows that $s = \tilde
s$, which proves the theorem. Obviously, the size of the trellis state set $S$
equals $p^\gamma$.
\eind
\pfend




\begin{thebibliography}{10}

\bibitem{artin91}
M.~Artin.
\newblock {\em Algebra}.
\newblock Prentice Hall, 1991.

\bibitem{chenCL94}
C-J. Chen, T-Y. Chen, and H-A. Loeliger.
\newblock Construction of linear ring codes for 6 {P}{S}{K}.
\newblock {\em IEEE Trans. Inf. Th.}, 40:563--566, 1994.

\bibitem{fagnaniZ96}
F.~Fagnani and S.~Zampieri.
\newblock Dynamical systems and convolutional codes over finite abelian groups.
\newblock {\em IEEE Trans. Inf. Th}, 42:1892--1912, 1996.

\bibitem{fagnaniZ01}
F.~Fagnani and S.~Zampieri.
\newblock System-theoretic properties of convolutional codes over rings.
\newblock {\em IEEE Trans. Inf. Th}, 47:2256--2274, 2001.

\bibitem{fornasiniP04}
E.~Fornasini and R.~Pinto.
\newblock Matrix fraction descriptions in convolutional coding.
\newblock {\em Linear Algebra and its Applications}, 392:119--158, 2004.

\bibitem{forneyT93}
G.D. Forney and M.D. Trott.
\newblock The dynamics of group codes: state spaces, trellis diagrams, and
  canonical encoders.
\newblock {\em IEEE Trans. Inf. Th}, 39:1491--1513, 1993.

\bibitem{gluesingS07}
H.~Gluesing-Luersen and G.~Schneider.
\newblock State space realizations and monomial equivalence for convolutional
  codes.
\newblock {\em Linear Algebra and its Applications}, 425:518--533, 2007.

\bibitem{johannessonW93}
R.~Johannesson and Z-X. Wan.
\newblock A linear algebra approach to minimal convolutional encoders.
\newblock {\em IEEE Trans. Inf. Th.}, IT-39:1219--1233, 1993.

\bibitem{johannessonWW98}
R.~Johannesson, Z-X. Wan, and E.~Wittenmark.
\newblock Some structural properties of convolutional codes over rings.
\newblock {\em IEEE Trans. Inf. Th.}, IT-44:839--845, 1998.

\bibitem{kailath}
T.~Kailath.
\newblock {\em Linear Systems}.
\newblock Prentice Hall, Englewood Cliffs, N.J, 1980.

\bibitem{kuijperPcastle08}
M.~Kuijper and R.~Pinto.
\newblock Minimal trellis construction for finite support convolutional ring
  codes.
\newblock In A.~Barbero, editor, {\em Coding Theory and Applications (ICMCTA)},
  LN in Computer Science 5228, pages 95--106. Springer, 2008.

\bibitem{kuijperPcontrolo08}
M.~Kuijper and R.~Pinto.
\newblock Minimal state diagrams for controllable behaviors over finite rings.
\newblock In {\em Proceedings of Controlo'08}, pages 1--6, Vila Real, Portugal,
  21-23 July 2008.

\bibitem{kuijperPP07}
M.~Kuijper, R.~Pinto, and J.~W. Polderman.
\newblock The predictable degree property and row reducedness for systems over
  a finite ring.
\newblock {\em Linear Algebra and its Applications}, 425:776--796, 2007.

\bibitem{kuijperS09}
M.~Kuijper and K.~Schindelar.
\newblock {G}r{\"{o}}bner bases and behaviors over finite rings.
\newblock submitted (March 2009) to 48th IEEE Conf. Decision and Control,
  Shanghai, China, 2009.

\bibitem{loeligerFMT94}
H-A Loeliger, G.D. Forney, T.~Mittelholzer, and M.D. Trott.
\newblock Minimality and observability of group systems.
\newblock {\em Linear Algebra and its Applications}, 205-206:937--963, 1994.

\bibitem{loeligerM96}
H-A Loeliger and T.~Mittelholzer.
\newblock Convolutional codes over groups.
\newblock {\em IEEE Trans. Inf. Th.}, IT-42:1660--1686, 1996.

\bibitem{masseyM89}
J.L. Massey and T.~Mittelholzer.
\newblock Convolutional codes over rings.
\newblock In {\em Proc. 4th Joint Swedish-Soviet Int. Workshop Information
  Theory}, pages 14--18, 1989.

\bibitem{masseyM90}
J.L. Massey and T.~Mittelholzer.
\newblock Systematicity and rotational invariance of convolutional codes over
  rings.
\newblock In {\em Proc. 2nd Int. Workshop on Algebraic and Combinatorial Coding
  Theory}, pages 154--159, 1990.

\bibitem{mcdonald74}
B.R. McDonald.
\newblock {\em Finite rings with identity}.
\newblock Marcel Dekker, New York, 1974.

\bibitem{mceliece98}
R.J. McEliece.
\newblock The algebraic theory of convolutional codes.
\newblock In R.A.~Brualdi V.S.~Pless, W.C.~Huffman, editor, {\em Handbook of
  Coding Theory Vol. 1}. North-Holland, Amsterdam, 1998.

\bibitem{mittelholzer93}
T.~Mittelholzer.
\newblock Minimal encoders for convolutional codes over rings.
\newblock In B.~Honory, M.~Darnell, and P.G. Farell, editors, {\em
  Communications Theory and Applications}, pages 30--36. HW Comm. Ltd, 1993.

\bibitem{rosenthalSY96}
J.~Rosenthal, J.M. Schumacher, and E.V. York.
\newblock On behaviors and convolutional codes.
\newblock {\em IEEE Trans. Inf. Th.}, 42:1881--1891, 1996.

\bibitem{soleS07}
P.~Sol\'{e} and V.~Sison.
\newblock Quaternary convolutional codes from linear block codes over galois
  rings.
\newblock {\em IEEE Trans. Inf. Th}, 53:2267--2270, 2007.

\bibitem{vaziraniSR96}
V.V. Vazirani, H.~Saran, and B.S. Rajan.
\newblock An efficient algorithm for constructing minimal trellises for codes
  over finite abelian groups.
\newblock {\em IEEE Trans. Inf. Th.}, 42:1839--1854, 1996.

\bibitem{wil88}
J.C. Willems.
\newblock Models for dynamics.
\newblock {\em Dynamics Rep.}, 2:171--282, 1988.

\bibitem{wittenmark98}
E.~Wittenmark.
\newblock {\em An Encounter with Convolutional Codes over Rings}.
\newblock PhD dissertation, Lund University, Lund, Sweden, 1998.

\bibitem{wittenmarkisit98}
E.~Wittenmark.
\newblock Minimal trellises for convolutional codes over rings.
\newblock In {\em Proceedings 1998 IEEE International Symposium in Information
  Theory}, page~15, Cambridge, USA, 1998.

\end{thebibliography}
\end{document}